\documentclass[10pt,twocolumn,prl,aps,amssymb,amsmath,tightenlines,showpacs]{revtex4}

\newcommand{\cC}{\ensuremath{\mathcal{C}}}
\newcommand{\cP}{\ensuremath{\mathcal{P}}}
\newcommand{\cT}{\ensuremath{\mathcal{T}}}
\newcommand{\half}{\mbox{$\textstyle{\frac{1}{2}}$}}
\newcommand{\quat}{\mbox{$\textstyle{\frac{1}{4}}$}}

\begin{document}

\title{Faster than Hermitian Quantum Mechanics}

\author{Carl~M.~Bender${}^{1,2}$, Dorje~C.~Brody${}^2$,
Hugh~F.~Jones${}^2$, and Bernhard~K.~Meister${}^{2,3}$}

\affiliation{${}^1$Physics Department, Washington University, St.
Louis, MO 63130, USA\\ ${}^2$Blackett Laboratory, Imperial College,
London SW7 2BZ, UK\\ ${}^3$Department of Physics, Renmin University
of China, Beijing 100872, China}

\date{\today}

\begin{abstract}
Given an initial quantum state $|\psi_I\rangle$ and a final quantum
state $|\psi_F\rangle$ in a Hilbert space, there exist Hamiltonians
$H$ under which $|\psi_I\rangle$ evolves into $|\psi_F\rangle$.
Consider the following quantum brachistochrone problem: Subject to
the constraint that the difference between the largest and smallest
eigenvalues of $H$ is held fixed, which $H$ achieves this
transformation in the least time $\tau$? For Hermitian Hamiltonians
$\tau$ has a nonzero lower bound. However, among non-Hermitian
$\cP\cT$-symmetric Hamiltonians satisfying the same energy
constraint, $\tau$ can be made arbitrarily small without violating
the time-energy uncertainty principle. This is because for such
Hamiltonians the path from $|\psi_I\rangle$ to $|\psi_F\rangle$ can
be made short. The mechanism described here is similar to that in
general relativity in which the distance between two space-time
points can be made small if they are connected by a wormhole. This
result may have applications in quantum computing.
\end{abstract}

\pacs{11.30.Er, 03.65.Ca, 03.65.Xp}

\maketitle

Suppose that one wishes to transform unitarily a state
$|\psi_I\rangle$ in a Hilbert space to a different state
$|\psi_F\rangle$ by means of a Hamiltonian $H$. In Hermitian quantum
mechanics, such a transformation requires a nonzero amount of time,
provided that the difference between the largest and the smallest
eigenvalues of $H$ is held fixed. However, if we extend quantum
mechanics into the complex domain while keeping the energy
eigenvalues real, then under the same energy constraint it is
possible to achieve such a transformation in an \emph{arbitrarily
short time}. In this paper we demonstrate this by means of simple
examples.

The paper is organized as follows: We first review why in Hermitian
quantum mechanics there is an unavoidable lower bound $\tau$ on the
time required to transform one state into another. In particular, we
consider the minimum time required to flip unitarily a spin-up state
into a spin-down state. We then summarize briefly how Hermitian
quantum mechanics can be extended into the complex domain while
retaining the reality of the energy eigenvalues, the unitarity of
time evolution, and the probabilistic interpretation. In this
complex framework we show how a spin-up state can be transformed
arbitrarily quickly to a spin-down state by a simple non-Hermitian
Hamiltonian. Then we discuss the transformation between pairs of
states by more general complex non-Hermitian Hamiltonians. We make
some comments regarding possible experimental consequences of these
ideas.

In Hermitian quantum mechanics how does one achieve the
transformation
$|\psi_I\rangle\to|\psi_F\rangle=e^{-iHt/\hbar}|\psi_I\rangle$ in
the shortest time $t=\tau$? Since $\tau$ is the minimum of all
possible evolution times $t$, the Hamiltonian associated with $\tau$
is the ``quantum brachistochrone'' \cite{re1}. Finding the optimal
evolution time requires only the solution to a much simpler problem,
namely, finding the optimal evolution time for the $2\times2$ matrix
Hamiltonians acting in the two-dimensional subspace spanned by
$|\psi_I\rangle$ and $|\psi_F\rangle$ \cite{re2}.

To solve the Hermitian version of the two-dimensional quantum
brachistochrone problem one can choose the basis so that the initial
and final states are given by
\begin{eqnarray}
|\psi_I\rangle = \left( \begin{array}{c} 1\\0\end{array}
\right)\quad{\rm and} \qquad |\psi_F\rangle=\left(\begin{array}{c}
a\\b\end{array}\right),\label{eq1}
\end{eqnarray}
with $|a|^2+|b|^2=1$. The most general $2\times2$ Hermitian
Hamiltonian has the form
\begin{eqnarray}
H = \left(\begin{array}{cc} s & r{\rm e}^{-i\theta} \cr r{\rm
e}^{i\theta} & u \end{array}\right), \label{eq2}
\end{eqnarray}
where the four parameters $r$, $s$, $u$, and $\theta$ are real. The
eigenvalue constraint $E_+-E_-=\omega$ reads
\begin{equation}
\omega^2 = (s-u)^2+4r^2. \label{eq3}
\end{equation}
The Hamiltonian $H$ in (\ref{eq2}) can be expressed in terms of the
Pauli matrices as $H=\half(s+u){\bf 1} + \half \omega
{\boldsymbol{\sigma}} \!\cdot\!{\bf n}$, where ${\bf
n}=\frac{2}{\omega} \left(r\cos\theta, r\sin\theta,
\frac{s-u}{2}\right)$ is a unit vector. Using the matrix identity
$\exp(i\phi\,{\boldsymbol{\sigma}}\!\cdot\!{\bf n})=\cos\phi\,{\bf
1} + i \sin\phi\,{\boldsymbol{\sigma}}\!\cdot\!{\bf n}$, the
relation $|\psi_F\rangle=e^{-iH\tau/\hbar}|\psi_I\rangle$ takes the
form
\begin{eqnarray}
\left( \begin{array}{c} a\\b\end{array} \right)
=e^{-\frac{1}{2}i(s+u)t/\hbar} \left( \begin{array}{c}
\cos\left(\frac{\omega t}{2\hbar}\right)-i\frac{s-u}
{\omega}\sin\left(\frac{\omega t}{2\hbar}\right) \cr {} \cr
-i\frac{2r}{\omega}e^{i\theta}\sin\left( \frac{\omega
t}{2\hbar}\right) \end{array} \right). \label{eq5}
\end{eqnarray}

From the second component of (\ref{eq5}) we obtain
$|b|=\frac{2r}{\omega}\sin \left( \frac{\omega t}{2\hbar}\right)$,
which gives the time required to transform the initial state:
$t=\frac{2\hbar}{\omega} \arcsin\big(\frac{\omega|b|}{2r}\big)$. We
optimize this relation over all $r>0$, keeping in mind that
(\ref{eq3}) gives a maximum value of $\half\omega$ for $r$, achieved
when $s=u$. The optimal time is thus
\begin{equation}
\tau = \frac{2\hbar}{\omega} \arcsin|b|. \label{eq6}
\end{equation}
Note that if $a=0$ and $b=1$ we have $\tau=2\pi\hbar/\omega$ for the
smallest time required to transform $\left(1\atop0\right)$ to the
orthogonal state $\left(0\atop1\right)$. This value of $\tau$ is
called the {\it passage time} \cite{re3}.

For general $a$ and $b$, at the optimal time $\tau$ we have
$a=e^{-i\tau s/\hbar}\sqrt{1-|b|^2}$ and $b=-ie^{-i\tau
s/\hbar}|b|e^{i\theta}$, which satisfies the condition
$|a|^2+|b|^2=1$ that the norm of the state does not change under
unitary time evolution. The parameters $s$ and $\theta$ are
determined by the phases of $a$ and $b$. Writing $a=|a|e^{i{\rm
arg}(a)}$ and $b=|b|e^{i{\rm arg}(b)}$, we find that the optimal
Hamiltonian is
\begin{eqnarray}
H = \left(\begin{array}{cc} \frac{\omega {\rm arg}(a)}{2\arcsin|b|}
& \frac{\omega}{4}e^{-i[{\rm arg}(b) -{\rm arg}(a)-\frac{\pi}{2}]}
\cr \frac{\omega}{4}e^{i[{\rm arg}(b) -{\rm arg}(a)-\frac{\pi}{2}] }
& \frac{\omega {\rm arg}(a)}{2\arcsin|b|} \end{array}\right).
\label{eq7}
\end{eqnarray}
Since the overall phase of $|\psi_{\rm F}\rangle$ is not physically
relevant, the quantity ${\rm arg}(a)$, for example, may be chosen
arbitrarily and without loss of generality we may assume that it is
$0$. We are free to choose ${\rm arg}(a)$ because there is no
absolute energy in quantum mechanics; one can add a constant to the
eigenvalues of the Hamiltonian without altering the physics.
Equivalently, this means that the value of $\tau$ cannot depend on
the trace $s+u$ of $H$.

Interpreting the result for $\tau$ in (\ref{eq6}) requires care
because while this equation resembles the time-energy uncertainty
principle, it is merely the statement that {\it
rate$\times$time$=$distance}. The constraint (\ref{eq3}) on $H$ is
equivalent to placing a bound on the standard deviation $\Delta H$
of the Hamiltonian, where $\Delta H$ in a normalized state
$|\psi\rangle$ is given by $(\Delta H)^2=\langle\psi|H^2
|\psi\rangle-\langle\psi|H|\psi\rangle^2$. The maximum value of
$\Delta H$ is $\omega/2$. According to the Anandan-Aharonov
relation~\cite{re4}, the speed of evolution of a quantum state is
given by $\Delta H$. The distance between the initial state
$|\psi_I\rangle$ and the final state $|\psi_F\rangle$ is
$\delta=2\arccos(|\langle \psi_F|\psi_I \rangle|)$. Thus, the
shortest time $\tau$ to achieve the evolution from $|\psi_I\rangle$
to $|\psi_F\rangle=e^{-iH\tau/\hbar} |\psi_I\rangle$ is bounded
below because the speed is bounded above while the distance is held
fixed. The Hamiltonian $H$ that realizes the shortest time evolution
can be understood as follows: The standard deviation $\Delta H$ of
the Hamiltonian in (\ref{eq2}) is $r$. Since $\Delta H$ is bounded
by $\omega/2$, to maximize the speed of evolution (and minimize the
time of evolution) we choose $r=\omega /2$.

The objective of this paper is to perform the same optimization for
complex non-Hermitian Hamiltonians having $\cP\cT$ symmetry. There
are infinitely many $\cP\cT$-symmetric complex non-Hermitian
Hamiltonians whose eigenvalues are real and bounded below. Here,
$\cP$ is the parity reflection operator and $\cT$ is the time
reversal operator. A one-parameter family of such Hamiltonians that
has been investigated intensively \cite{re5,re6} is given by
$H=p^2+x^2(ix)^\epsilon$, where $\epsilon>0$. Although this
Hamiltonian is not Hermitian in the usual Dirac sense, where
Hermitian adjoint consists of complex conjugation and matrix
transposition, $H$ defines a unitary theory of quantum mechanics
\cite{re7,re8}. This is because $H$ is self-adjoint with respect to
a new inner product that is different from the Dirac inner product
of conventional quantum mechanics.

This new inner product is expressed in terms of a linear operator
$\cC$ that satisfies three equations \cite{re9}:
\begin{equation}
\cC^2=1,\quad [\cC,H]=0,\quad {\rm and}\quad [\cC,\cP\cT]=0.
\label{eq8}
\end{equation}
For any given $H$ we can, in principle, calculate $\cC$ by solving
the three equations in (\ref{eq8}). We then define an inner product
in terms of $\cC\cP\cT$ conjugation. Thus, in a
$\mathcal{PT}$-symmetric quantum theory the inner product is
specified dynamically in terms of the Hamiltonian. Furthermore, the
time-evolution operator $e^{-iHt/\hbar}$ is unitary (norm
preserving) because $H$ commutes with $\cC\cP\cT$.

It has been shown that for any $\cP\cT$-symmetric Hamiltonian having
real eigenvalues there exists an equivalent Hermitian Hamiltonian
\cite{re10}. The argument goes as follows: The Hermitian operator
$\cC\cP$ is positive and can thus be written as $\cC\cP= e^Q$. By
means of the similarity transformation $\tilde H=e^{-Q/2}He^{Q/2}$
one can construct the corresponding Hermitian Hamiltonian $\tilde
H$, which is equivalent to $H$ in the sense that it has the same
eigenvalues. The states in the $\cP\cT$-symmetric theory are mapped
by the operator $e^{-Q/2}$ to corresponding states in the Hermitian
theory. But, since this operator does not keep the states in the
same Hilbert space, relative properties of states can be changed.
For example, the overlap distance between two states does not remain
constant in the original Hilbert space. In this paper we exploit
this property of the $e^{-Q/2}$ transformation to circumvent the
Hermitian limit on $\tau$.

We now show how to solve the $\mathcal{PT}$-symmetric
brachistochrone problem for a simple non-Hermitian two-dimensional
matrix Hamiltonian of the form
\begin{eqnarray}
H=\left(\begin{array}{cc} r e^{i\theta}&s\cr s & r e^{-i\theta}
\end{array}\right). \label{eq9}
\end{eqnarray}
(This Hamiltonian was examined in detail in Ref.~\cite{re7}.) To
show that $H$ in (\ref{eq9}) is $\mathcal{PT}$ symmetric, we let
$\cal{T}$ be the operation of complex conjugation and $\cal{P}$ be
given by
\begin{eqnarray}
\cal{P}=\left(\begin{array}{cc} 0&1\cr 1 & 0\end{array}\right).
\label{eq10}
\end{eqnarray}
The eigenvalues $E_\pm =r\cos\theta \pm \sqrt{s^2 -r^2
\sin^2\theta}$ of $H$ in (\ref{eq9}) are real provided that $s^2>r^2
\sin^2\theta$. This inequality defines the region of unbroken
$\cP\cT$ symmetry.

The unnormalized eigenstates of $H$ are
\begin{eqnarray}
|E_+\rangle = \left( \begin{array}{c} e^{i\alpha/2} \cr e^{-i
\alpha/2} \end{array} \right),\quad |E_-\rangle = \left(
\begin{array}{c} i e^{-i \alpha/2} \cr -i e^{i \alpha/2}\end{array}
\right), \label{eq12}
\end{eqnarray}
where the real parameter $\alpha$ is defined by
$\sin\alpha=(r/s)\sin\theta$. The operator $\cC$
satisfying the conditions in (\ref{eq8}) is given by
\begin{eqnarray}
{\cal C} = \frac{1}{\cos\alpha} \left(
\begin{array}{cc} i \sin\alpha & 1 \cr 1 & -i\sin\alpha \end{array}
\right). \label{eq13}
\end{eqnarray}
By using (\ref{eq10}) and (\ref{eq13}) we calculate that the
$\cal{CPT}$ norms of both eigenstates in (\ref{eq12}) are
$\sqrt{2\cos\alpha}$.

Following the procedure used for Hermitian Hamiltonians, we rewrite
$H$ in (\ref{eq9}) in the form $H=(r\cos\theta){\bf 1}+\half \omega
{\boldsymbol{\sigma}}\!\cdot\!{\bf n}$, where ${\bf
n}=\frac{2}{\omega}(s,0,ir\sin\theta)$ is a unit vector and the
squared difference between the energy eigenvalues is
\begin{eqnarray}
\omega^2=4s^2-4r^2\sin^2\theta. \label{eq15}
\end{eqnarray}
The positivity of $\omega^2$ is ensured by the condition of unbroken
$\cP\cT$ symmetry. This equation emphasizes the key difference
between Hermitian and $\cP\cT$-symmetric Hamiltonians: The
corresponding equation (\ref{eq3}) for a Hermitian matrix
Hamiltonian has a \emph{sum} of squares, while this equation for
$\omega^2$ has a \emph{difference} of squares. Thus, Hermitian
Hamiltonians exhibit elliptic behavior, which leads to a nonzero
lower bound for the optimal time $\tau$. The hyperbolic nature of
(\ref{eq15}) allows $\tau$ to approach zero because, as we will see,
the matrix elements of the $\cP\cT$-symmetric Hamiltonian can be
made large without violating the energy constraint $E_+-E_-=\omega$.

The $\cP\cT$-symmetric analog of the evolution equation
(\ref{eq5}) is given by
\begin{eqnarray}
e^{-iHt/\hbar}\left( \begin{array}{c} 1\\0\end{array} \right) =
\frac{e^{-itr\cos\theta/\hbar}}{\cos\alpha} \left( \begin{array}{c}
\cos\left(\frac{\omega t}{2\hbar}-\alpha\right) \\ {}\\
-i\sin\left( \frac{\omega t}{2\hbar}\right) \end{array} \right).
\label{eq16}
\end{eqnarray}
We apply this result to the pair of vectors examined in the
Hermitian case: $|\psi_I\rangle =\left(1\atop0\right)$ and
$|\psi_F\rangle=\left(0\atop1\right)$. (Note that these vectors are
not orthogonal with respect to the $\cC\cP\cT$ inner product.) From
(\ref{eq16}) we see that the evolution time to reach
$|\psi_F\rangle$ from $|\psi_I\rangle$ is
$t=(2\alpha+\pi)\hbar/\omega$. Optimizing this result over allowable
values for $\alpha$, we see that as $\alpha$ approaches $-\half\pi$
the optimal time $\tau$ tends to zero.

Note that in the limit $\alpha\to-\half\pi$ we get $\cos\alpha\to0$.
However, in terms of the variable $\alpha$ the energy constraint
(\ref{eq15}) becomes $\omega^2=4s^2\cos^2 \alpha$. Since $\omega$ is
held fixed, in order to have $\alpha$ approach $-\frac{1}{2}\pi$ we
must require that $s\gg1$. It then follows from the relation
$\sin\alpha=(r/s)\sin\theta$ that $|r|\sim|s|$, so we must also
require that $r\gg1$. Evidently, in order to make $\tau\ll1$ the
matrix elements of the $\cP\cT$-symmetric Hamiltonian (\ref{eq9})
must be large.

The result demonstrated here does not violate the uncertainty
principle. Indeed, a Hermitian Hamiltonian and a non-Hermitian
$\cP\cT$-symmetric Hamiltonian both share the properties that (i)
the passage time is given by $2\pi\hbar/\omega$, and (ii) $\Delta
H\leq\omega/2$. The key difference is that a pair of states such as
$\left(1\atop0\right)$ and $\left(0\atop1\right)$ are orthogonal in
a Hermitian theory but have the separation $\delta=\pi-2|\alpha|$ in
the $\cP\cT$-symmetric theory. This is because the Hilbert space
metric of the $\cP\cT$-symmetric quantum theory depends on the
Hamiltonian. As a consequence, it is possible to set the parameter
$\alpha$ to create a wormhole-like effect in the Hilbert space
\cite{re11}.

A {\it gedanken} experiment to realize this effect in a laboratory
might work as follows: We use a Stern-Gerlach filter to create a
beam of spin-up electrons. The beam then passes through a `black
box' containing a device governed by a $\cP\cT$-symmetric
Hamiltonian that flips the spins unitarily in a very short time. The
outgoing beam then enters a second Stern-Gerlach device that
verifies that the electrons are now in spin-down states. In effect,
the black-box device is applying a magnetic field in the complex
direction $(s,0, ir\sin\theta)$. If the field strength is
sufficiently strong, then spins can be flipped unitarily in
virtually no time because the complex path joining these two states
is arbitrary short without violating the energy constraint. The
arbitrarily short alternative complex pathway from an up state to a
down state, as illustrated by this thought experiment, is
reminiscent of the short alternative distance between two widely
separated space-time points as measured through a wormhole in
general relativity \cite{re12}.

The $\cP\cT$-symmetric Hamiltonian (\ref{eq9}) used in the foregoing
illustrative example contains only three arbitrary real parameters,
which are not sufficient to allow the initial state
$|\psi_I\rangle=\left(1\atop0\right)$ to evolve into any final state
$|\psi_F\rangle$. Indeed, it follows from (\ref{eq16}) that
$\left(1\atop0\right)$ can only evolve into $\left(a\atop b\right)$
if the relative phase of $a$ and $b$ is $\pm\half\pi$. Therefore, we
introduce the more general four-real-parameter $\cP\cT$-symmetric
Hamiltonian
\begin{eqnarray}
H=\left( \begin{array}{cc} x+(z+iy) & \frac{z}{\tan\gamma}
-iy\tan\gamma \\ \frac{z}{\tan\gamma} -iy \tan\gamma &
x-(z+iy)\end{array} \right), \label{eq17}
\end{eqnarray}
which is associated with a more general definition of parity
reflection $\cP$ than that used in (\ref{eq10}):
\begin{eqnarray}
\cal{P}=\left(\begin{array}{cc} \sin\gamma & \cos\gamma \cr
\cos\gamma & -\sin\gamma \end{array}\right). \label{eq18}
\end{eqnarray}
We retain the same definition
for $\cT$, namely, that $\cT$ performs complex conjugation.

As before, we express $H$ in the form $H=x{\bf 1}+
\frac{\omega}{2}{\boldsymbol{\sigma}}\!\cdot\!{\bf n}$, where in
this case ${\bf n}=\frac{2}{\omega}
(z/\tan\gamma-iy\tan\gamma,0,z+iy)$. The operator ${\cal
C}={\boldsymbol{\sigma}}\cdot{\bf n}$ is given by
\begin{eqnarray}
\mathcal{C}= \frac{2}{\omega}\left(\begin{array}{cc} z+iy &
\frac{z}{\tan\gamma} -iy \tan\gamma \\ \frac{z}{\tan\gamma} -iy
\tan\gamma & -z-iy \end{array} \right), \label{eq19}
\end{eqnarray}
which, along with (\ref{eq18}), allows us to define the inner
product with respect to which the Hamiltonian (\ref{eq17}) becomes
self-adjoint. The energy constraint $E_+-E_-=\omega$ again takes a
hyperbolic form:
\begin{eqnarray}
\omega^2 = 4z^2{\rm csc}^2\gamma - 4y^2\sec^2\gamma.
\label{eq20}
\end{eqnarray}

For this more general Hamiltonian, the initial state
$\left(1\atop0\right)$ evolves as follows:
\begin{eqnarray}
e^{-iHt/\hbar}\left(\begin{array}{c}1\\0\end{array}\right) &=&
e^{-ixt/\hbar} \nonumber \\ && \hspace{-2.2cm} \times \left(
\begin{array}{c} \cos\left(\frac{\omega t} {2\hbar}\right) +
\frac{2y}{\omega} \sin\left(\frac{\omega t}{2\hbar}\right) -
i\frac{2z}{\omega} \sin\left(\frac{\omega t} {2\hbar}\right) \\
{}\vspace{-0.2cm} \\ -\frac{2y}{\omega}\tan\gamma \sin\left(
\frac{\omega t}{2\hbar}\right) - i \frac{2z} {\omega\tan\gamma}
\sin\left(\frac{\omega t}{2\hbar}\right)\end{array}\right).
\label{eq21}
\end{eqnarray}
The time evolution preserves the $\cC\cP\cT$ norm of the initial
state, which is $\left(\frac{2z}{\omega\sin\gamma}\right)^{1/2}$. We
therefore choose the general form of the final state to be
\begin{equation}
|\psi_F\rangle=\sqrt{{\textstyle\frac{2z}{\omega\sin\gamma}}}
\left(\begin{array}{c}ue^{iA}\\ve^{i(A+\xi)}\end{array}\right),
\label{eq22}
\end{equation}
where $u$, $v$, $A$, and $\xi$ are real parameters.

We now introduce dimensionless variables $X=2x/\omega$,
$Y=2y/\omega$, $Z=2z/\omega\sin\gamma$, and $T=\omega t/2\hbar$, as
well as the shifted variable $B=A+xt/\hbar$. By identifying the
right sides of (\ref{eq21}) and (\ref{eq22}), we obtain
\begin{eqnarray}
\begin{array}{l} \cos T +Y \sin T = \sqrt{Z} u \cos B,\\
Y\tan\gamma \sin T =-\sqrt{Z} v\cos(B+\xi), \\
\sqrt{Z} \sin\gamma \sin T =-u \sin B, \\
\sqrt{Z}\cos\gamma \sin T =-v\sin(B+\xi). \end{array}
\label{eq23}
\end{eqnarray}
The energy constraint in (\ref{eq20}) takes the hyperbolic form
\begin{equation}
1=Z^2-Y^2\sec^2\gamma. \label{eq24}
\end{equation}
The condition that the norm of the initial vector be preserved
under time evolution imposes the requirement that $u^2+v^2+ 2uvY
\sin\xi/ Z\cos\gamma=1$, which can be derived from the five
equations in (\ref{eq23}) and (\ref{eq24}).

The generic problem is now to pick a final vector; that is, to
choose the parameters $A$, $u$, $v$, and $\xi$ in (\ref{eq22}), and
then to solve (\ref{eq23}) and (\ref{eq24}) to determine the
parameters $X$, $Y$, $Z$, and $\gamma$ for which $|\psi_I\rangle$
reaches $|\psi_F\rangle$ under the Hamiltonian in (\ref{eq17}). We
then must find the smallest value of $T$ for which the
transformation is realized.

To illustrate the procedure, we consider the example $u=v$ and
solve the five simultaneous equations in the form of Laurent
series valid for small $T$. The result is:
\begin{eqnarray}
\begin{array}{l}
X = \left( \quat\pi-\half\xi -A \right)T^{-1} + {\rm O}(T), \\
Y =- uT^{-1}\cos(\quat\pi-\half\xi)  + {\rm O}(T), \\
Z =- uT^{-1} + {\rm O}(T), \\
\gamma = \quat\pi-\half\xi+{\rm O}(T^2). \end{array} \label{eq25}
\end{eqnarray}
Note that the parameter $T$ may be taken arbitrarily small and thus
the initial vector evolves into the final vector in an optimal time
that is arbitrarily small. Of course, the matrix elements of the
Hamiltonian become large in this limit, but this is possible because
the energy constraint in (\ref{eq20}) is hyperbolic in character.

We conclude by remarking that the results established here provide
the possibility of performing experiments that definitively
distinguish between Hermitian and $\mathcal{PT}$-symmetric
Hamiltonians. If practical implementation of complex
$\mathcal{PT}$-symmetric Hamiltonians were proved feasible, then the
identification of the optimal unitary transformation would be
particularly important in the design and implementation of fast
quantum communication and computation algorithms (cf. \cite{re13}).

Of course, the wormhole-like effect we have discussed here can only
be realized if we can switch rapidly between Hermitian and
$\mathcal{PT}$-symmetric Hamiltonians by means of similarity
transformations. It is conceivable that there is a sort of quantum
protection mechanism that places a lower bound on the time required
to switch Hilbert spaces. If so, this would limit the applicability
of a Hilbert-space wormhole to improve quantum algorithms.

\vskip1pc We thank D. W. Hook for useful discussions. CMB thanks the
U.S.~Department of Energy and DCB thanks the The Royal Society for
support.

\end{document}